\documentclass[12pt]{article}
\def\be{\begin{equation}}
\def\ee{\end{equation}}
\def\ber{\begin{eqnarray}}
\def\eer{\end{eqnarray}}
\usepackage{amsfonts,amssymb}
\begin{document}
\vspace*{1cm}
\begin{center}
{\Large \bf Charmed and Charmed-Strange Mesons\\[1ex]  in Kaluza-Klein Picture}\\

\vspace{4mm}

{\large A.A. Arkhipov\\
{\it State Research Center ``Institute for High Energy Physics" \\
 142280 Protvino, Moscow Region, Russia}}\\
\end{center}

\vspace{4mm}
\begin{abstract}
{In the present paper, we continue our study the structure of
hadronic spectra, started in Refs. \cite{1,2,3,4,5}, from the view
point of existence the extra dimensions in the spirit of Kaluza-Klein
approach. We show that all charmed and charmed-strange mesons,
including the recently observed new states \cite{7,8}, are
excellently incorporated in the systematics provided by Kaluza-Klein
approach.}
\end{abstract}

\section{Introduction}

In the present paper, we continue our study the structure of hadronic
spectra, started in Refs. \cite{1,2,3,4,5}, from the view point of
existence the extra dimensions in the spirit of Kaluza-Klein
approach. Here we shall concern the spectra of charmed and
charmed-strange mesons. The physics of charmed and charmed-strange
mesons is really charmed and charmed-strange one. On the one hand the
fundamental QCD Lagrangian predicted the charmed and charmed-strange
states containing $c$ and $s$ quarks, while on the other hand there
are  serious problems to describe that sort of experimentally
observed states in the framework of known QCD-inspired potential
models with the quark and gluon degrees of freedom. The best
currently performed lattice computations in QCD \cite{9} can not help
us to understand the exact nature of the recently observed new states
in the spectrum of charmed-strange mesons too. However, we show below
that all charmed and charmed-strange mesons, including the recently
observed new states, are excellently incorporated in the systematics
provided by Kaluza-Klein approach.

\section{Charmed mesons}

As it was established in previous paper \cite{5}, the charmed
$D^0$(1864)-meson may occupy $M_{19}^{2K}(1860-1864)$-Storey in
Kaluza-Klein tower of KK excitations in two-kaon system and
$M_{21}^{2K}(2003-2006)$-Storey in the same tower is acceptable for
the $D^{*0}$(2006)-meson. However, $D^{*0}$(2006)-meson has been
observed as a resonance in $D^0\pi^0$--system. That is why, first of
all, let us build the Kaluza-Klein tower of KK excitations for
$D\pi$--system by the formula
\begin{equation}\label{Dpi}
M_n^{D\pi} = \sqrt{m_{D}^2+\frac{n^2}{R^2}} +
\sqrt{m_{\pi}^2+\frac{n^2}{R^2}},\quad (n=1,2,3,\ldots),
\end{equation}
where $R$ is the fundamental scale characterizing the size of extra
dimensions calculated early from the analysis of nucleon-nucleon
dynamics at low energies \cite{1,2}
\begin{equation}\label{scale}
\frac{1}{R} = 41.481\,\mbox{MeV}\quad \mbox{or}\quad
R=24.1\,\mbox{GeV}^{-1}=4.75\,10^{-13}\mbox{cm}.
\end{equation}
The Kaluza-Klein tower such built is shown in Table 1 where the
comparison with experimentally observed mass spectrum of
$D^{*}$-mesons is also presented.

Throughout we have used Review of Particle Physics \cite{6} where the
experimental data on mass spectrum of the resonance states have been
extracted from. In particular, we have used $m_{D^0}=1864.1\pm
1.0$\,Mev and $m_{D^\pm}=1869.4\pm 0.5$\,Mev for the masses of
$D$-mesons calculating the Kaluza-Klein tower for $D\pi$-system. In
Table 2 -- Table 3 we collected some known experimental information.
The Tables~1--3 show a remarkable correspondence of the calculated KK
excitations for $D\pi$ system with the experimentally measured masses
of the $D^*$-mesons. In fact, there are many empty cells in Table 1
where we have not found the corresponding experimental data.

At the next step we built the Kaluza-Klein tower of KK excitations
for the $D^*\pi$-system by the formula
\begin{equation}\label{Dxpi}
M_n^{D^*\pi} = \sqrt{m_{D^*}^2+\frac{n^2}{R^2}} +
\sqrt{m_\pi^2+\frac{n^2}{R^2}},\quad (n=1,2,3,\ldots),
\end{equation}
and this is shown in Table 4 where the comparison with experimentally
observed mass spectrum is also presented. Some known experimental
information in that case is collected in separate tables: Table 5 --
Table 6. We have used $m_{D^{*0}}=2006.71\pm 0.5$\,Mev and
$m_{D^{*\pm}}=2010.0\pm 0.5$\,Mev for the masses of $D^*$-mesons
calculating the Kaluza-Klein tower for $D^*\pi$-system. Again we see
from Tables 4--6 that there is a remarkable correspondence of the
calculated KK excitations for $D^*\pi$-system with the experimentally
measured masses of the resonance states. Here, there are many empty
cells in Table 4 as well, where we have not found the corresponding
experimental data. However, it should be noted that the decay modes
$D_1^0(2420)\rightarrow D^{*+}(2010)\pi^-$ and
$D_1^+(2420)\rightarrow D^{*0}(2007)\pi^+$ have been seen but the
decay modes $D_1^0(2420)\rightarrow D^{+}\pi^-$ and
$D_1^+(2420)\rightarrow D^{0}\pi^+$ have not been observed \cite{6}.
Table 1 and Table 4, as it were, confirm that observation.

\section{Charmed-strange mesons}

Now, we go to the charmed-strange mesons. In the first, we calculate
the Kaluza-Klein tower of KK excitations for the
$K^{*0}K^{*\pm}$-system by the formula
\begin{equation}\label{2Kx}
M_n^{K^{*0}K^{*\pm}} = \sqrt{m_{K^{*0}}^2+\frac{n^2}{R^2}} +
\sqrt{m_{K^{*\pm}}^2+\frac{n^2}{R^2}},\quad (n=1,2,3,\ldots),
\end{equation}
which is shown in Table 7. We see that $D_s^\pm(1969)$-meson lives in
$M_{10}^{K^{*0}K^{*\pm}}$-Storey of that Kaluza-Klein tower. We have
used $m_{K^{*0}}=896.1\pm 0.27$\,Mev and $m_{K^{*\pm}}=891.66\pm
0.26$\,Mev for the masses of $K^*(892)$-mesons calculating the
Kaluza-Klein tower. It should be emphasized that
$M_{9}^{K^{*0}K^{*\pm}}$-Storey in this Kaluza-Klein tower is quite
acceptable for the $f_2(0^+2^{++})(1950)$-meson (LASS 91)\cite{6}. In
Table 8 we presented some known experimental information concerning
$D_s^\pm$-meson.

We farther present the results calculating the Kaluza-Klein tower of
KK excitations for the $DK$-system by the formula
\begin{equation}
M_n^{DK} = \sqrt{m_D^2+\frac{n^2}{R^2}} +
\sqrt{m_K^2+\frac{n^2}{R^2}},\quad (n=1,2,3,\ldots).\label{DK}
\end{equation}
These results are shown in Table 9. We have used $m_{D^0}=1864.1\pm
1.0$\,Mev and $m_{D^{\pm}}=1869.4\pm 0.5$\,Mev for the masses of
$D$-mesons calculating the Kaluza-Klein tower. Table 10  concerns the
experimental data of resonance states in $DK$-system extracted from
\cite{6}. Recently CLEO Collaboration reported the observation of a
narrow resonance $D_{sJ}^+(2463)$ in $D_s^{*+}\pi^0$-system \cite{7}.
From Table 9 we see that $M_7^{DK}(2459-2468)$-Storey is acceptable
for the resonance in $D^{+}K^0$-system with a similar mass.

We have also calculated the Kaluza-Klein tower of KK excitations for
the $D^*K$-system by the formula
\begin{equation}
M_n^{D^*K} = \sqrt{m_{D^*}^2+\frac{n^2}{R^2}} +
\sqrt{m_K^2+\frac{n^2}{R^2}},\quad (n=1,2,3,\ldots).\label{DxK}
\end{equation}
This is shown in Table 11. Here, we have used $m_{D^{*0}}=2006.7\pm
0.5$\,Mev and $m_{D^{*\pm}}=2010.0\pm 0.5$\,Mev for the masses of
$D^*$-mesons calculating the Kaluza-Klein tower. As it is seen, the
$D_{s1}^\pm(2536)$-meson is excellently incorporated in Table 11. We
have also found that $D_{sJ}^\pm(2573)$-meson may occupy the
$M_6^{D^*K}$-Storey even though the decay $D_{sJ}^+(2573)\rightarrow
D^{*0(+)}(2007)K^{+(0)}$ has not been seen so far. Table 12 contains
the experimental data in respect of $D_{s1}^\pm(2536)$-meson.

In Table 13 we present the results calculating the Kaluza-Klein tower
of KK excitations for the $D_s^\pm(1969)\pi$-system by the formula
\begin{equation}
M_n^{D_s^\pm\pi} = \sqrt{m_{D_s^\pm}^2+\frac{n^2}{R^2}} +
\sqrt{m_\pi^2+\frac{n^2}{R^2}},\quad (n=1,2,3,\ldots),\label{Dspi}
\end{equation}
where it follows that the $D_{sJ}^\pm(2112)$ lives in the first
Storey of this Kaluza-Klein tower from. The seventh Storey of this
Kaluza-Klein tower is acceptable for the recently discovered in
\cite{8} and confirmed in \cite{7} $D_{sJ}(2317)$-meson as well.

At the same time we have calculated Kaluza-Klein tower of KK
excitations for the $D_s^{*\pm}(2112)\pi$-system by the formula
\begin{equation}
M_n^{D_{sJ}\pi} = \sqrt{m_{D_s^{*\pm}}^2+\frac{n^2}{R^2}} +
\sqrt{m_\pi^2+\frac{n^2}{R^2}},\quad (n=1,2,3,\ldots),\label{Dsxpi}
\end{equation}
and found that the recently discovered narrow resonance of mass 2.46
GeV decaying to $D_s^{*+}\pi^0$ \cite{7} lives in the seventh Storey
of this Kaluza-Klein tower: see Table 14.

At last, we present in Table 15 the results calculating the
Kaluza-Klein tower of KK excitations for the $K^*(892)K$-system by
the formula
\begin{equation}\label{KxK}
M_n^{K^*K} = \sqrt{m_{K^*}^2+\frac{n^2}{R^2}} +
\sqrt{m_K^2+\frac{n^2}{R^2}},\quad (n=1,2,3,\ldots).
\end{equation}
Here we have found that the recently discovered in \cite{8} and
confirmed in \cite{7} $D_{sJ}(2317)$-meson \cite{8} may excellently
be incorporated in $M_{22}^{K^*K}(2315-2317)$-Storey of that
Kaluza-Klein tower. Moreover, the first Storey in this tower is
acceptable for the $h_1(?^-1^{+-})(1380)$-meson with experimentally
measured mass $1386\pm19$ MeV ($\Gamma=91\pm30$) MeV(AVERAGE
PDG)\cite{6}, and $M_{4}^{K^*K}(1432)$-Storey is very acceptable for
the $f_1(0^+1^{++})(1420)$-meson with experimentally measured mass
$1433.4\pm0.8$ MeV ($\Gamma=58.8\pm3.3$) MeV(SPEC 99)\cite{6} as
well, $M_{6}^{K^*K}(1482)$-Storey is a good place for the
$\eta(0^+0^{-+})(1440)$-meson with experimentally measured mass
$1475\pm5$ MeV ($\Gamma=81\pm11$) MeV(AVERAGE PDG)\cite{6}, the
$f_1(0^+1^{++})(1510)$-meson with experimentally measured mass
$1512\pm4$ MeV ($\Gamma=35\pm15$) MeV(MPS 88)\cite{6} may all rights
to occupy $M_{7}^{K^*K}(1514)$-Storey, and finally
$M_{10}^{K^*K}(1632)$-Storey is very acceptable for the
$\eta_2(0^+2^{-+})(1645)$-meson with experimentally measured mass
$1632\pm14$ MeV ($\Gamma=180^{\,+\,22}_{\,-\,20}$) MeV(AVERAGE
PDG)\cite{6} too.

\section{Summary}

We calculated the Kaluza-Klein towers of KK excitations for the
different experimentally observed charmed and charmed-strange
hadronic systems and found that all known charmed and charmed-strange
mesons, including the recently observed new states, are excellently
incorporated in the systematics provided by Kaluza-Klein picture.
This is a very non-trivial fact, even though there are many empty
cells in Tables 1,4,7,9,11, 13,14,15 where we have no the
corresponding experimental information.

Of course, it would be very desirable to state new experiments to
search new states, and we believe that the Tables presented here  may
serve as a guide for the physicists--experimenters.


\newpage

\begin{center}
Table 1: Kaluza-Klein tower of KK excitations for $D\pi$-system and
experimental data.

\vspace{5mm} {\large
\begin{tabular}{|c|c|c|c|c|c|}\hline
 n & $ M_n^{D^0 \pi^0}$MeV & $ M_n^{D^ 0 \pi^{\pm}}$MeV & $
 M_n^{D^\pm \pi^0}$MeV & $ M_n^{D^\pm \pi^\pm}$MeV & $
 M_{exp}^{D \pi}$\,MeV  \\
 \hline
1  & 2005.77 & 2010.17 & 2011.07 & 2015.46 & $D^{*0,\pm}(\frac{1}{2}1^-)$ \\
2  & 2024.38 & 2028.31 & 2029.67 & 2033.61 &  \\
3  & 2051.84 & 2055.24 & 2057.13 & 2060.53 &  \\
4  & 2085.36 & 2088.29 & 2090.64 & 2093.57 &  \\
5  & 2123.06 & 2125.60 & 2128.33 & 2130.86 &  \\
6  & 2163.77 & 2165.99 & 2169.03 & 2171.25 &  \\
7  & 2206.79 & 2208.75 & 2212.02 & 2213.99 &  \\
8  & 2251.66 & 2253.41 & 2256.88 & 2258.63 &  \\
9  & 2298.10 & 2299.68 & 2303.30 & 2304.88 &  \\
10 & 2345.92 & 2347.36 & 2351.09 & 2352.53 &  \\
11 & 2394.97 & 2396.30 & 2400.12 & 2401.44 &  \\
12 & 2445.17 & 2446.39 & 2450.29 & 2451.51 & $D_2^{*0,\pm}(\frac{1}{2}2^+)$ \\
13 & 2496.42 & 2497.56 & 2501.52 & 2502.65 &  \\
14 & 2548.68 & 2549.74 & 2553.74 & 2554.80 &  \\
15 & 2601.89 & 2602.88 & 2606.92 & 2607.91 &  \\
16 & 2656.01 & 2656.94 & 2661.01 & 2661.94 &  \\
17 & 2711.01 & 2711.89 & 2715.97 & 2716.84 &  \\
18 & 2766.84 & 2767.67 & 2771.76 & 2772.59 &  \\
19 & 2823.49 & 2824.27 & 2828.37 & 2829.16 &  \\
20 & 2880.91 & 2881.66 & 2885.76 & 2886.50 &  \\
21 & 2939.10 & 2939.81 & 2943.90 & 2944.61 &  \\
22 & 2998.01 & 2998.69 & 3002.77 & 3003.46 &  \\
23 & 3057.64 & 3058.29 & 3062.35 & 3063.01 &  \\
24 & 3117.95 & 3118.57 & 3122.62 & 3123.25 &  \\
25 & 3178.92 & 3179.52 & 3183.55 & 3184.16 &  \\
26 & 3240.54 & 3241.12 & 3245.13 & 3245.71 &  \\
27 & 3302.78 & 3303.34 & 3307.33 & 3307.89 &  \\
28 & 3365.63 & 3366.17 & 3370.13 & 3370.67 &  \\
29 & 3429.06 & 3429.58 & 3433.51 & 3434.03 &  \\
30 & 3493.05 & 3493.55 & 3497.46 & 3497.96 &  \\ \hline
\end{tabular}}
\end{center}

\newpage

\vspace*{10mm}
\begin{center}
Table 2: $M_{1}^{D\pi}(2006-2015)$--Storey.

\vspace{5mm}
\begin{tabular}{|c|c|c|c|c|}\hline
$R(IJ^{P})$ & $ M_R $\, MeV & $ \Gamma_R $\, MeV & Reaction & Collab.
\\ \hline
$D^{*0}(\frac{1}{2}1^-)$ & 2006.7 $\pm$ 0.5 & $< 2.1$ & AVERAGE & PDG
00 \\
$D^{*\pm}(\frac{1}{2}1^-)$ & 2010.0 $\pm$ 0.5 & $< 0.131$ & AVERAGE &
PDG 00 \\
\hline
\end{tabular}
\end{center}

\vspace{10mm}
\begin{center}
Table 3: $M_{12}^{D\pi}(2445-2452)$--Storey.

\vspace{5mm}
\begin{tabular}{|c|c|c|c|r|}\hline
$R(IJ^{P})$ & $ M_R $\, MeV & $ \Gamma_R $\, MeV & Reaction &
Collab.\ \ \ \\ \hline $D_2^{*0}(\frac{1}{2}2^+) $ & 2453$\pm$3$\pm$2
& 25$\pm$10$\pm$5& $\gamma Be
\rightarrow D^+\pi^-X$ & E687 94 \\
 & 2458.9 $\pm$ 2.0 & 23 $\pm$ 5 & AVERAGE & PDG 00 \\
$D_2^{*\pm}(\frac{1}{2}2^+) $ & 2453$\pm$3$\pm$2 & 23$\pm$9$\pm$5&
$\gamma Be
\rightarrow D^0\pi^+X$ & E687 94 \\
 & 2459 $\pm$ 4 & $25^{\,+\,8}_{\,-\,7}$ & AVERAGE & PDG 00 \\
\hline
\end{tabular}
\end{center}

\newpage

\begin{center}
Table 4: Kaluza-Klein tower of KK excitations in $D^*\pi$-system and
experimental data.

\vspace{5mm} {\large
\begin{tabular}{|c|c|c|c|c|c|}\hline
 n & $ M_n^{D^{*0} \pi^0}$MeV & $ M_n^{D^{*0} \pi^{\pm}}$MeV & $
 M_n^{D^{*\pm} \pi^0}$MeV & $ M_n^{D^{*\pm} \pi^\pm}$MeV & $
 M_{exp}^{D^*\pi}$\,MeV  \\
 \hline
1  & 2148.34 & 2152.73 & 2151.63 & 2156.03 &  \\
2  & 2166.85 & 2170.78 & 2170.15 & 2174.08 &  \\
3  & 2194.14 & 2197.55 & 2197.44 & 2200.84 &  \\
4  & 2227.44 & 2230.37 & 2230.73 & 2233.66 &  \\
5  & 2264.85 & 2267.38 & 2268.13 & 2270.67 &  \\
6  & 2305.21 & 2307.43 & 2308.48 & 2310.70 &  \\
7  & 2347.81 & 2349.77 & 2351.07 & 2353.04 &  \\
8  & 2392.20 & 2393.96 & 2395.46 & 2397.22 &  \\
9  & 2438.11 & 2439.70 & 2441.36 & 2442.94 & $D_1(2420)$ \\
10 & 2485.35 & 2486.79 & 2488.58 & 2490.02 &  \\
11 & 2533.76 & 2535.09 & 2536.98 & 2538.30 &  \\
12 & 2583.27 & 2584.49 & 2586.47 & 2587.69 &  \\
13 & 2633.79 & 2634.92 & 2636.97 & 2638.11 & $D^{*\pm}(2640)$ \\
14 & 2685.26 & 2686.32 & 2688.43 & 2689.49 &  \\
15 & 2737.64 & 2738.63 & 2740.79 & 2741.78 &  \\
16 & 2790.89 & 2791.82 & 2794.03 & 2794.96 &  \\
17 & 2844.98 & 2845.86 & 2848.10 & 2848.97 &  \\
18 & 2899.87 & 2900.70 & 2902.97 & 2903.80 &  \\
19 & 2955.54 & 2956.33 & 2958.62 & 2959.40 &  \\
20 & 3011.97 & 3012.72 & 3015.02 & 3015.77 &  \\
21 & 3069.12 & 3069.83 & 3072.15 & 3072.86 &  \\
22 & 3126.98 & 3127.66 & 3129.98 & 3130.67 &  \\
23 & 3185.53 & 3186.18 & 3188.51 & 3189.16 &  \\
24 & 3244.74 & 3245.37 & 3247.70 & 3248.32 &  \\
25 & 3304.60 & 3305.20 & 3307.53 & 3308.14 &  \\
26 & 3365.09 & 3365.67 & 3367.10 & 3368.58 &  \\
27 & 3426.19 & 3426.75 & 3429.07 & 3429.63 &  \\
28 & 3487.88 & 3488.42 & 3490.74 & 3491.28 &  \\
29 & 3550.15 & 3550.67 & 3552.98 & 3553.50 &  \\
30 & 3612.98 & 3613.48 & 3615.78 & 3616.29 &  \\ \hline
\end{tabular}}
\end{center}

\newpage

\begin{center}
Table 5: $M_{9}^{D^*\pi}(2438-2443)$--Storey.

\vspace{3mm}
\begin{tabular}{|c|c|c|c|r|}\hline
$R(IJ^{P})$ & $ M_R $\, MeV & $ \Gamma_R $\, MeV & Reaction &
Collab.\ \ \\ \hline $D_1^{0}(\frac{1}{2}1^{+})$ & 2428$\pm$3$\pm$2 &
$23^{\,+\,8\,+\,10}_{\,-\,6\,-\,3}$ &
$e^+e^- \rightarrow D^{*+}\pi^-X$ & CLEO 90 \\
 & 2428$\pm$8$\pm$5 & $58 \pm 14 \pm 10$ & $\gamma N \rightarrow D^{*+}\pi^-X$ & TPS 89 \\
 & 2422.2 $\pm$ 1.8 & $18.9^{\,+\,4.6}_{\,-\,3.5}$ & AVERAGE & PDG 00 \\
$D_1^{\pm}(\frac{1}{2}?^{?})$ & $2425 \pm 2 \pm 2$ &
$26^{\,+\,8}_{\,-\,7}\pm 4$ &
$e^+e^- \rightarrow D^{*0}\pi^+X$ & CLE2 94 \\
 & $2443 \pm 7 \pm 5$ & $41 \pm 19 \pm 8$ & $\gamma N \rightarrow D^{*0}\pi^+X$ & TPS 89 \\
 & 2427 $\pm$ 5 & $28 \pm 8$ & AVERAGE & PDG 00 \\
\hline
\end{tabular}
\end{center}

\vspace{3mm}
\begin{center}
Table 6: $M_{13}^{D^*\pi}(2634-2638)$--Storey.

\vspace{3mm}
\begin{tabular}{|c|c|c|c|r|}\hline
$R(IJ^{P})$ & $ M_R $\, MeV & $ \Gamma_R $\, MeV & Reaction &
Collab.\ \ \\ \hline $D^{*\pm}(\frac{1}{2}?^{?})$ & $2637 \pm 2 \pm
6$ & $<$ 15 & $e^+e^-
\rightarrow D^{*+} \pi^+\pi^-X $ & DLPH 98 \\
\hline
\end{tabular}
\end{center}

\newpage

\begin{center}
Table 7: Kaluza-Klein tower of KK excitations in
$K^{*0}K^{*\pm}$-system and $D_s^\pm$-meson.

\vspace{5mm} {\large
\begin{tabular}{|c|c|c|c|c|c|}\hline
 n & $ M_n^{K^{*0}K^{*\pm}}$MeV & $
 M_{exp}^{K^{*0}K^{*\pm}}$\,MeV  \\
 \hline
1  & 1789.68 &  \\
2  & 1795.44 &  \\
3  & 1805.00 &  \\
4  & 1818.30 &  \\
5  & 1835.25 &  \\
6  & 1855.77 &  \\
7  & 1879.72 &  \\
8  & 1906.98 &  \\
9  & 1937.42 & $f_2(0^+2^{++})$ \\
10 & 1970.88 & $D_s^\pm(00^-)$ \\
11 & 2007.21 &  \\
12 & 2046.27 &  \\
13 & 2087.89 &  \\
14 & 2131.93 &  \\
15 & 2178.24 &  \\
16 & 2226.67 &  \\
17 & 2277.12 &  \\
18 & 2329.40 &  \\
19 & 2383.44 &  \\
20 & 2439.10 &  \\
21 & 2496.28 &  \\
22 & 2554.87 &  \\
23 & 2614.78 &  \\
24 & 2675.92 &  \\
25 & 2738.22 &  \\
26 & 2801.58 &  \\
27 & 2865.94 &  \\
28 & 2931.24 &  \\
29 & 2997.42 &  \\
30 & 3064.41 &  \\ \hline
\end{tabular}}
\end{center}


\vspace{3mm}
\begin{center}
Table 8: $M_{10}^{K^{*0}K^{*\pm}}(1971)$--Storey.

\vspace{5mm}
\begin{tabular}{|c|c|c|c|r|}\hline
$R(IJ^{P})$ & $ M_R $\, MeV & $ \Gamma_R $\, MeV & Reaction &
Collab.\ \ \\ \hline $D_s^\pm(00^-)$ & $1970\pm 5\pm 5$  &  &
$e^+e^-$ & CLEO 83 \\
 & $1969.0\pm 1.4$  &  & AVERAGE & PDG 00 \\
\hline
\end{tabular}
\end{center}

\newpage

\begin{center}
Table 9: Kaluza-Klein tower of KK excitations in $DK$-system and
experimental data.

\vspace{5mm} {\large
\begin{tabular}{|c|c|c|c|}\hline
 n & $ M_n^{D^\pm K^0}$\,MeV & $ M_n^{D^0K^\pm}$\,MeV & $
 M_{exp}^{DK}$\,MeV   \\
 \hline
1  & 2369.26 & 2359.98 &  \\
2  & 2375.78 & 2366.54 &  \\
3  & 2386.53 & 2377.37 &  \\
4  & 2401.35 & 2392.28 &  \\
5  & 2420.03 & 2411.08 &  \\
6  & 2442.33 & 2433.51 &  \\
7  & 2468.00 & 2459.32 & $D_{sJ}$(2463)? \\
8  & 2496.79 & 2488.25 &  \\
9  & 2528.45 & 2520.06 &  \\
10 & 2562.75 & 2554.51 & $D_{sJ}^\pm$(2573) \\
11 & 2599.47 & 2591.38 &  \\
12 & 2638.42 & 2630.48 &  \\
13 & 2679.43 & 2671.64 &  \\
14 & 2722.34 & 2714.68 &  \\
15 & 2767.00 & 2759.48 &  \\
16 & 2813.29 & 2805.90 &  \\
17 & 2861.09 & 2853.84 &  \\
18 & 2910.32 & 2903.19 &  \\
19 & 2960.87 & 2953.86 &  \\
20 & 3012.67 & 3005.78 &  \\
21 & 3065.64 & 3058.86 &  \\
22 & 3119.73 & 3113.06 &  \\
23 & 3174.85 & 3168.29 &  \\
24 & 3230.98 & 3224.52 &  \\
25 & 3288.04 & 3281.69 &  \\
26 & 3346.00 & 3339.75 &  \\
27 & 3404.82 & 3398.65 &  \\
28 & 3464.44 & 3458.37 &  \\
29 & 3524.84 & 3518.87 &  \\
30 & 3585.99 & 3580.10 &  \\ \hline
\end{tabular}}
\end{center}

\vspace{5mm}
\begin{center}
Table 10: $M_{10}^{DK}(2555-2563)$--Storey.

\vspace{5mm}
\begin{tabular}{|c|c|c|c|r|}\hline
$R(IJ^{P})$ & $ M_R $\, MeV & $ \Gamma_R $\, MeV & Reaction &
Collab.\ \ \\ \hline $D_{sJ}^\pm(0?^?)$ & 2573.5 $\pm$ 1.7 &
$15^{\,+\,5}_{\,-\,4}$ & AVERAGE & PDG 00 \\
\hline
\end{tabular}
\end{center}

\newpage

\begin{center}
Table 11: Kaluza-Klein tower of KK excitations in $D^*K$-system and
experimental data.

\vspace{5mm} {\large
\begin{tabular}{|c|c|c|c|}\hline
 n & $ M_n^{D^{*\pm} K^0}$MeV & $ M_n^{D^{*0} K^{\pm}}$MeV & $ M_{exp}^{D^{*}K}$\,MeV  \\
 \hline
1  & 2509.83 & 2502.55 &  \\
2  & 2516.25 & 2509.01 &  \\
3  & 2526.84 & 2519.67 &  \\
4  & 2541.44 & 2534.36 & $D_{s1}^\pm$(2536) \\
5  & 2559.83 & 2552.87 &  \\
6  & 2581.79 & 2574.94 & $D_{sJ}^\pm$(2573)? \\
7  & 2607.05 & 2600.34 &  \\
8  & 2635.38 & 2628.80 &  \\
9  & 2666.51 & 2660.08 &  \\
10 & 2700.24 & 2693.94 &  \\
11 & 2736.33 & 2730.17 &  \\
12 & 2774.61 & 2768.58 &  \\
13 & 2814.89 & 2809.00 &  \\
14 & 2857.02 & 2851.26 &  \\
15 & 2900.87 & 2895.23 &  \\
16 & 2946.31 & 2940.78 &  \\
17 & 2993.22 & 2987.81 &  \\
18 & 3041.52 & 3036.22 &  \\
19 & 3091.12 & 3085.92 &  \\
20 & 3141.93 & 3136.83 &  \\
21 & 3193.89 & 3188.89 &  \\
22 & 3246.94 & 3242.02 &  \\
23 & 3301.01 & 3296.18 &  \\
24 & 3356.05 & 3351.31 &  \\
25 & 3412.02 & 3407.36 &  \\
26 & 3468.87 & 3464.29 &  \\
27 & 3526.56 & 3522.06 &  \\
28 & 3585.05 & 3580.63 &  \\
29 & 3644.31 & 3639.96 &  \\
30 & 3704.31 & 3700.03 &  \\ \hline
\end{tabular}}
\end{center}

\newpage

\vspace*{10mm}
\begin{center}
Table 12: $M_{4}^{D^*K}(2534-2541)$--Storey.

\vspace{5mm}
\begin{tabular}{|c|c|c|c|r|}\hline
$R(IJ^{P})$ & $ M_R $\, MeV & $ \Gamma_R $\, MeV & Reaction &
Collab.\ \ \\ \hline $D_{s1}^\pm(01^{+})$ & $2536.6\pm 0.7\pm 0.4 $ &
$<$ 5.44 & $e^+e^- \rightarrow
D^{*+}K^0X $ & CLEO 90 \\
 & $2535.2\pm 0.5\pm 1.5 $ & $<$ 3.9 & $e^+e^- \rightarrow
D^{*0}K^+X $ & ARG 92 \\
 & $2534.8\pm 0.6\pm 0.6 $ & $<$ 2.3 & $e^+e^- \rightarrow
D^{*+}K^0X $ & CLE2 93 \\
 & $2535.3\pm 0.2\pm 0.5 $ & $<$ 2.3 & $e^+e^- \rightarrow
D^{*0}K^+X $ & CLE2 93 \\
 & $2535\pm 0.6\pm 1 $ & $<$ 3.2 & $\gamma Be \rightarrow
D^{*0,+}K^{+,\,0}X $ & E687 94 \\
 & $2535.35\pm 0.34 $ & $<$ 2.3 & AVERAGE & PDG 00 \\
\hline
\end{tabular}
\end{center}

\newpage

\begin{center}
Table 13: Kaluza-Klein tower of KK excitations in $D_s^\pm\pi$-system
and experimental data.

\vspace{5mm} {\large
\begin{tabular}{|c|c|c|c|c|c|}\hline
 n & $ M_n^{D_s^\pm \pi^0}$MeV & $ M_n^{D_s^\pm \pi^{\pm}}$MeV & $
 M_{exp}^{D_s^\pm \pi}$\,MeV   \\
 \hline
1  & 2110.64 & 2115.04 & $D_s^{*\pm}$(2112) \\
2  & 2129.18 & 2133.11 &    \\
3  & 2156.52 & 2159.92 &    \\
4  & 2189.87 & 2192.80 &    \\
5  & 2227.35 & 2229.89 &    \\
6  & 2267.80 & 2270.02 &    \\
7  & 2310.50 & 2312.47 & $D_{sJ}$(2317)   \\
8  & 2355.02 & 2356.77 &    \\
9  & 2401.06 & 2402.65 &    \\
10 & 2448.44 & 2449.88 &    \\
11 & 2497.02 & 2498.34 &    \\
12 & 2546.70 & 2547.92 &    \\
13 & 2597.40 & 2598.53 &    \\
14 & 2649.07 & 2650.13 &    \\
15 & 2701.66 & 2702.65 &    \\
16 & 2755.14 & 2756.07 &    \\
17 & 2809.45 & 2810.33 &    \\
18 & 2864.58 & 2865.41 &    \\
19 & 2920.50 & 2921.29 &    \\
20 & 2977.17 & 2977.92 &    \\
21 & 3034.59 & 3035.30 &    \\
22 & 3092.72 & 3093.40 &    \\
23 & 3151.54 & 3152.19 &    \\
24 & 3211.03 & 3211.66 &    \\
25 & 3271.18 & 3271.78 &    \\
26 & 3331.95 & 3332.53 &    \\
27 & 3393.34 & 3393.90 &    \\
28 & 3455.33 & 3455.87 &    \\
29 & 3517.90 & 3518.42 &    \\
30 & 3581.02 & 3581.53 &    \\ \hline
\end{tabular}}
\end{center}

\newpage

\begin{center}
Table 14: Kaluza-Klein tower of KK excitations in
$D_s^{*\pm}\pi$-system and $D_{sJ}^+$(2463)-meson.

\vspace{5mm} {\large
\begin{tabular}{|c|c|c|c|c|c|}\hline
 n & $ M_n^{D_s^{*\pm} \pi^0}$MeV & $ M_n^{D_s^{*\pm} \pi^{\pm}}$MeV & $
 M_{exp}^{D_s^{*\pm} \pi}$\,MeV  \\
 \hline
1  & 2254.01 & 2258.41 &    \\
2  & 2272.46 & 2276.39 &    \\
3  & 2299.65 & 2303.05 &    \\
4  & 2332.80 & 2335.73 &    \\
5  & 2370.02 & 2372.55 &    \\
6  & 2410.14 & 2412.36 &    \\
7  & 2452.47 & 2454.43 & $D_{sJ}^+$(2463) \\
8  & 2496.56 & 2498.31 &    \\
9  & 2542.12 & 2543.70 &    \\
10 & 2588.96 & 2590.41 &    \\
11 & 2636.96 & 2638.28 &    \\
12 & 2686.01 & 2687.23 &    \\
13 & 2736.04 & 2737.17 &    \\
14 & 2786.99 & 2788.05 &    \\
15 & 2838.82 & 2839.81 &    \\
16 & 2891.50 & 2892.43 &    \\
17 & 2944.98 & 2945.86 &    \\
18 & 2999.24 & 3000.07 &    \\
19 & 3054.26 & 3055.05 &    \\
20 & 3110.01 & 3110.76 &    \\
21 & 3166.46 & 3167.18 &    \\
22 & 3223.61 & 3224.30 &    \\
23 & 3281.43 & 3282.08 &    \\
24 & 3339.90 & 3340.53 &    \\
25 & 3399.00 & 3399.61 &    \\
26 & 3458.72 & 3459.30 &    \\
27 & 3519.04 & 3519.60 &    \\
28 & 3579.95 & 3580.49 &    \\
29 & 3641.42 & 3641.94 &    \\
30 & 3703.44 & 3703.94 &    \\ \hline
\end{tabular}}
\end{center}

\newpage

\begin{center}
Table 15: Kaluza-Klein tower of KK excitations in $K^*K$-system and
$D_{sJ}$(2317)-meson.

\vspace{5mm} {\large
\begin{tabular}{|c|c|c|c|c|c|}\hline
 n & $ M_n^{K^{*\pm} K^0}$MeV & $ M_n^{K^{*0} K^{\pm}}$MeV & $
 M_{exp}^{K^{*}K}$\,MeV  \\
 \hline
1  & 1392.02 & 1392.48 & $h_1(?1^{+-})$ \\
2  & 1400.05 & 1400.53 &   \\
3  & 1413.30 & 1413.82 &   \\
4  & 1431.57 & 1432.15 & $f_1(0^+1^{++})$ \\
5  & 1454.63 & 1455.27 &   \\
6  & 1482.18 & 1482.89 & $\eta(0^+0^{-+})$ \\
7  & 1513.94 & 1514.71 & $f_1(0^+1^{++})$ \\
8  & 1549.58 & 1550.42 &   \\
9  & 1588.80 & 1589.70 &   \\
10 & 1631.30 & 1632.27 & $\eta_2(0^+2^{-+})$ \\
11 & 1676.82 & 1677.83 &   \\
12 & 1725.08 & 1726.14 &   \\
13 & 1775.85 & 1776.95 &   \\
14 & 1828.91 & 1830.04 &   \\
15 & 1884.06 & 1885.22 &   \\
16 & 1941.12 & 1942.29 &   \\
17 & 1999.92 & 2001.11 &   \\
18 & 2060.32 & 2061.51 &   \\
19 & 2122.17 & 2123.38 &   \\
20 & 2185.37 & 2186.57 &   \\
21 & 2249.79 & 2251.00 &   \\
22 & 2315.35 & 2316.55 & $D_{sJ}$(2317) \\
23 & 2381.94 & 2383.14 &   \\
24 & 2449.49 & 2450.68 &   \\
25 & 2517.92 & 2519.10 &   \\
26 & 2587.17 & 2588.34 &   \\
27 & 2657.17 & 2658.33 &   \\
28 & 2727.87 & 2729.01 &   \\
29 & 2799.22 & 2800.35 &   \\
30 & 2871.17 & 2872.28 &   \\ \hline
\end{tabular}}
\end{center}
\end{document}